# Statistical study on propagation characteristics of Omega signals (VLF) in magnetosphere detected by the Akebono satellite

I Made Agus Dwi Suarjaya[1,2], Yoshiya Kasahara[1*] and Yoshitaka Goto[1]

**Abstract**

This paper shows a statistical analysis of 10.2 kHz Omega broadcasts of an artificial signal broadcast from ground stations, propagated in the plasmasphere, and detected using an automatic detection method we developed. We study the propagation patterns of the Omega signals to understand the propagation characteristics that are strongly affected by plasmaspheric electron density and the ambient magnetic field. We show the unique propagation patterns of the Omega 10.2 kHz signal when it was broadcast from two high–middle-latitude stations. We use about eight years of data captured by the Poynting flux analyzer subsystem on board the Akebono satellite from October 1989 to September 1997. We demonstrate that the signals broadcast from almost the same latitude (in geomagnetic coordinates) propagated differently depending on the geographic latitude. We also study propagation characteristics as a function of local time, season, and solar activity. The Omega signal tended to propagate farther on the nightside than on the dayside and was more widely distributed during winter than during summer. When solar activity was at maximum, the Omega signal propagated at a lower intensity level. In contrast, when solar activity was at minimum, the Omega signal propagated at a higher intensity and farther from the transmitter station.

## Introduction

The Earth's plasmasphere is located in the inner part of the magnetosphere and is mainly filled with cold plasma. Very low frequency (VLF) waves such as whistlers (Carpenter 1963) and Omega signals propagate in the magnetosphere as whistler mode waves and are strongly affected by the electron density profile. The major species of ions in the plasmasphere are protons, helium ions, and oxygen ions. The composition ratio of these ions plays an important role in the propagation effect of VLF waves such as the subprotonospheric whistler and the magnestospherically reflected whistler (Kimura 1966). Hence, it is very important to clarify the spatial distribution of electron density as well as the ion constituents in the magnetosphere to understand the propagation characteristics of VLF waves. In other words, the propagation characteristics of the whistler mode wave in the VLF range could be an important clue to study the electron density profile in the magnetosphere and the impact of these waves on the wider system.

Several studies had previously been conducted to determine the electron density profile around the Earth. Some studies of in situ satellite observations revealed features of the plasmasphere such as "notches" (Sandel et al. 2001, 2003). A particularly long-lived (2–3 days) depleted region [magnetic local time (MLT) >1.5–2 h) or "notch" extended out from $L \sim 3$ in the plasmasphere (Kotova et al. 2004). Remote sensing from ground-based magnetometers enabled the comparison of mass density and electron density between different $L$-shells using multiple observation stations along the 330° magnetic

*Correspondence: kasahara@is.t.kanazawa-u.ac.jp
[1] Graduate School of Natural Science and Technology, Kanazawa University, Kakuma-machi, Kanazawa 920-1192, Japan
Full list of author information is available at the end of the article longitude, spanning the $L$-shell between 1.5 and 3.4 (Chi et al. 2013). It was also demonstrated that mass density decreased within a few hours by 50% or more at 0.5 radius of the Earth or further inward of the plasmapause during moderately disturbed periods on December 10–15, 2003 (Menk et al. 2014). Measurement of electromagnetic wave data is also important to determine the electron density. For example, dispersion of electromagnetic waves originating from lighting discharge, known as whistlers (Helliwell 1965), is useful for estimating the spatial electron density profile in the plasmasphere, because the propagation velocity of whistler mode waves depends on the electron density profile around the Earth (Crouchley 1964; Singh et al. 2003). Statistical studies of whistlers have been performed using ground observatory data at Tihany, Hungary (Tarcsai et al. 1988; Collier et al. 2006; Lichtenberger et al. 2008). Another statistical study of whistlers was carried out using the data from Akebono satellite (Oike et al. 2014). These studies have provided information on the absorption effect of whistlers in the ionosphere. A recent statistical study by Zheng et al. (2016) also reported that one-to-one coincidences between whistlers observed by Van Allen Probes (Radiation Belt Storm Probes) and lightning strokes detected by World Wide Lightning Location Network are clearly shown in the $L$-shell range of $L = 1$–$3$.

Because the electron density profile around the Earth changes day by day, it is important to assess the trend of this change using statistical study of electromagnetic wave propagation over several years. The intensity of whistlers, however, depends on many factors such as lightning current, location, orientation, and ionospheric attenuation. On the other hand, signals observed from artificial transmitters are worth analyzing for such purposes because these transmitters continuously broadcast signals with constant power. For this reason, it is possible to analyze the extent of change in the wave propagation. This technique has been used to estimate the ionospheric topside and bottom-side profiles with sounders (Reinisch and Huang 2001; Ganguly et al. 2001), which are capable of checking the ionospheric peak parameters such as density, height, and peak shape (Pulinets et al. 2001).

Japan launched a satellite nicknamed Akebono (EXOS-D) in 1989 to observe the Earth's magnetosphere and plasmasphere. The satellite consists of several scientific instruments (i.e., particle detectors, magnetometer, electric field detector, plasma wave instruments, and auroral camera). The VLF instrument is one of the instruments on board Akebono used to measure VLF plasma waves, and the Poynting flux analyzer (PFX) is a subsystem of the VLF instrument (Kimura et al. 1990). The PFX is a waveform receiver that measures three components of the magnetic field and two components of the electric field. The waveforms generated from the PFX have bandwidth of 50 Hz at fixed center frequency between 100 Hz and 12.75 kHz. The wide dynamic range amplifier (WIDA) hybrid integrated circuits (ICs) are equipped for gain control to achieve wide measurable dynamic range (Kimura et al. 1990).

The Omega system was a navigation system used to provide a navigational aid not only for domestic aviation but also for oceanic air and marine shipping, and had eight ground station transmitters located in Norway, Liberia, Hawaii, North Dakota, La Reunion Island, Argentina, Australia, and Japan (Morris et al. 1994). The PFX observed Omega signals broadcast from these eight ground stations from 1989 to 1997. The PFX captured Omega signals at 10.2 kHz, which is one of the common frequencies among the transmission patterns of these eight stations. In 1997, this navigation system was shut down when replaced by the global positioning system.

Sawada et al. (1993) analyzed the Omega signals captured by the PFX and proposed a method to estimate global plasmaspheric electron density deduced from wave normal directions and in situ electron density along the Akebono trajectories. This method could derive a tomographic electron density profile comparing the wave normal direction and delay time (propagation time from transmission station to the observation point) of the Omega signals using 1-h data of a single satellite observation along a trajectory with the propagation paths calculated by a ray-tracing method (Kimura et al. 2001 and the references therein). The algorithm was further improved with a flexible method and novel stochastic algorithm (Goto et al. 2003). This enabled separate estimation of the effects of the ionosphere and plasmasphere.

Since the transmission pattern of frequency, transmission time, and location of each Omega station are known (described by Morris et al. 1994 and Suarjaya et al. 2016), it is easy to distinguish the signal source measured by the PFX. We can then determine many propagation properties such as attenuation ratio, propagation direction, delay time from the transmission station, and observation point along the satellite trajectories. Such parameters depend strongly on the plasma parameters along the propagation path. Therefore, it is valuable to analyze such propagation characteristics of VLF waves in the ionosphere and plasmasphere statistically using long-term observation data. In the present paper, we investigate statistical features of the Omega signals detected by the PFX from 1989 to 1997. To achieve this purpose, we extracted Omega signals from the enormous amount of PFX data using an automatic detection program developed by Suarjaya et al. (2016).

The outline of the present paper is as follows: In the next section, we briefly introduce our detection algorithm (Suarjaya et al. 2016) and the datasets used for the statistical study, in which we investigated the Omega signals broadcast from the high–middle-latitude stations in Norway and North Dakota. Next, we show the results of propagation characteristics depending on the MLT, season, and solar activity. Finally, discussion and conclusions of the present study are described.

## Data processing and automatic detection methods

As was also mentioned in the previous section, the Omega signal had been broadcast based on the specific pattern of frequency and time duration assigned to each transmission station. Each station had its own unique transmission pattern every 10 s with four common frequencies and one unique frequency. In the present paper, we analyzed the PFX data when the center frequency of the PFX was fixed at 10.2 kHz, which was one of the common frequencies of the Omega signal.

The PFX data are stored as waveforms using the common data format (CDF) file format. We have already developed a method in the Java programming language to detect the presence of Omega signals measured by the PFX and then analyze them comprehensively (Suarjaya et al. 2016). The flow of the data processing is described in the paper by Suarjaya et al. (2016) in more detail, but we briefly outline the process as follows.

In the automatic detection process, we first performed a FFT analysis of the PFX waveform data to detect the signal around 10.2 kHz. When the monochromatic signal at 10.2 kHz was detected in the PFX data, we determined the station that broadcast the signal, referring to the unique frequency pattern for each of the eight different Omega stations. For this process, it was first necessary to synchronize the Omega time with UTC time because the Omega system did not have leap-second correction. Next, we calculated the delay time of the signal by subtracting the raise time of the signal from the transmission time. The raise time is the arrival time of the Omega signal at 10.2 kHz, as determined by detecting the sudden increase in signal intensity at the center frequency (10.2 kHz), compared with those of surrounding frequencies (Suarjaya et al. 2016). As a next step, the signal intensity was determined when the average intensity at the center frequency was large enough to compare with those of surrounding frequency and ambient noise. Some error detection and correction were also applied to produce reliable results. The detection method developed by Suarjaya et al. (2016) is based on transmission frequency pattern, transmission time, and location of each Omega station combined with a fixed threshold, that minimizes the contamination of the VLF sources (for example, from other Omega stations). In addition, comparison of upper and lower frequencies introduced in their detection method enables exclusion of any events contaminated by whistlers.

## Data analysis

### Datasets

We first show the number of PFX data that we have processed, as shown in Fig. 1 and Table 1. Because we applied several kinds of operation modes to PFX in changing its center frequency, we selected the PFX data when its center frequency was fixed at 10.2 kHz to measure the Omega signal. In the figure and table, we count one event that corresponds to 10 s duration of data when the center frequency of the PFX was fixed at 10.2 kHz. From October 1989 to September 1997, we measured 3,030,614 events. The number of events changes every year. We obtained the highest number of events (408,790) in 1992 and the lowest number (213,155) in 1989. We also separated the datasets of events into four sides for the analysis based on magnetic local time (MLT). In the present study, we defined dawnside, dayside, duskside, and nightside as MLT from 3 to 9, from 9 to 15, from 15 to 21, and from 21 to 3, respectively. Comparing the number of events for each MLT or year, there were no significant differences except for the data in 1989. This is because this year only contains data for three months. Therefore, the datasets coverage for each year and each MLT are reliable enough for the statistical analyses in this study.

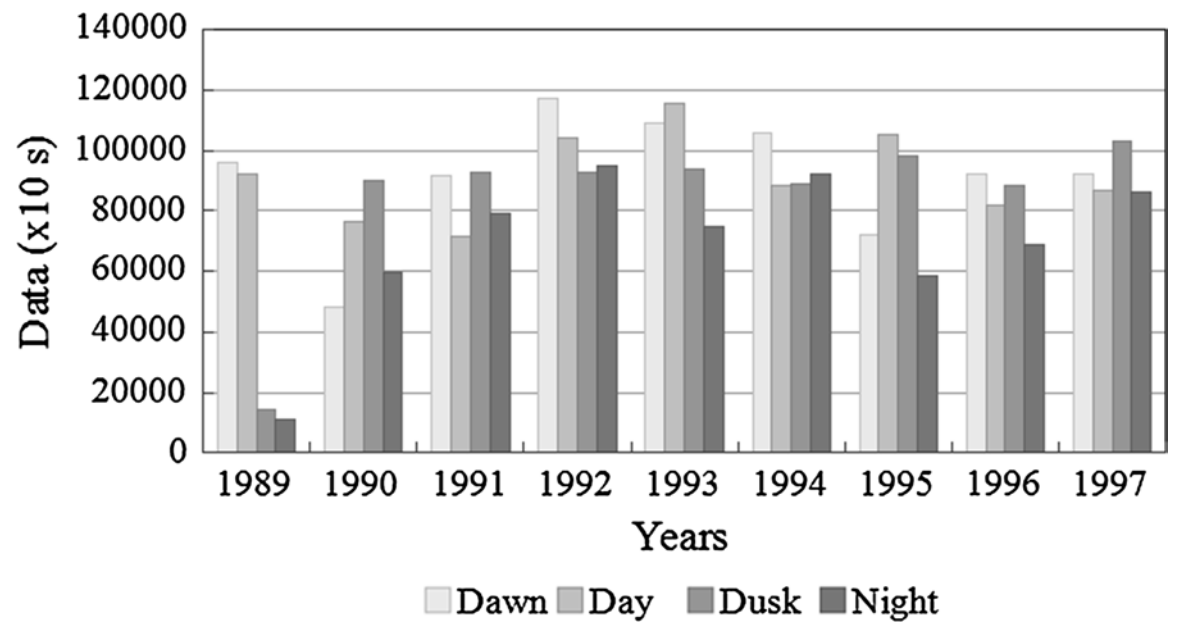

**Fig. 1** Number of data events processed from October 1989 to September 1997. Each event corresponds to 10 s duration of PFX data

**Table 1 Number of data events processed from October 1989 to September 1997**

| | Dawn | Day | Dusk | Night | Total |
|---|---|---|---|---|---|
| 1989 | 95,854 | 92,003 | 14,297 | 11,001 | 213,155 |
| 1990 | 48,231 | 76,448 | 89,670 | 59,446 | 273,795 |
| 1991 | 91,456 | 71,240 | 92,925 | 79,300 | 334,921 |
| 1992 | 117,240 | 104,322 | 92,652 | 94,576 | 408,790 |
| 1993 | 108,802 | 115,379 | 93,782 | 74,649 | 392,612 |
| 1994 | 105,783 | 88,317 | 88,951 | 91,894 | 374,945 |
| 1995 | 72,069 | 105,293 | 97,962 | 58,127 | 333,451 |
| 1996 | 92,192 | 81,758 | 88,379 | 68,660 | 330,989 |
| 1997 | 92,197 | 86,643 | 102,857 | 86,259 | 367,956 |
| Total | 823,824 | 821,403 | 761,475 | 623,912 | 3,030,614 |

Using these datasets, we comprehensively analyzed the Omega signals broadcast from each station. For the present paper, however, we analyzed only the signals broadcast from Norway and North Dakota stations located at high–middle-latitude regions because the orbital condition of Akebono was suitable to detect the signals from these stations along the ambient magnetic field lines due to its elliptical and high-inclination orbit. In the analyses, a dipole model was used for the geomagnetic field model and invariant latitude was referred from the "orbit datasets" provided by the Akebono project team. Regarding events in our analyses, we have "data" events and "detected" events, where a data event means data available whether detected or not, while detected events refer to events detected within the data of the Omega signal.

## General characteristics of Omega signal propagation from Norway and North Dakota Stations

First, we studied the signal intensity just above the transmission station to evaluate how widely and how far the Omega signals penetrated through the ionosphere over the station. For this purpose, we restricted the PFX data obtained below 640 km in altitude. Figure 2a shows the number of events from October 1989 to September 1997 used for the analysis just above the Norway station, which was located at latitude 55.96° N and longitude 100.72° E (geomagnetic coordinates). The number of events is indicated by the color scale given at the right of the panel, and white areas on the map indicate unavailability of PFX data. Figure 2b, c shows the averaged intensity of the signals in the magnetic and electric fields, respectively. Magnetic or electric signal intensity is indicated by the color scale given at the right of each panel. The gray color in Fig. 2b, c indicates that no Omega signal was detected, although the PFX was in operation and measuring at 10.2 kHz. This is because the intensity was too low compared to the threshold level for signal detection. It is noted that the threshold level is not a fixed absolute value but was adaptively determined according to the level of background noise. The detailed algorithm used to determine the threshold level is described in Suarjaya et al. (2016). We divided the vertical axis into 20 bins of latitude (5° for each bin) and the horizontal axis into 20 bins of longitude (5° for each bin). The location of the station is shown on the map (indicated by a rectangle in each panel). It is also notable that the high-intensity area was located ~10° north of the transmitter station. The center of the penetration region was located approximately at latitude 65° N and longitude 97° E (geomagnetic coordinates). The intensities at this point were approximately −265 dBT (magnetic field) and −120 dBV/m (electric field), where dBT and dBV/m are magnetic and electric

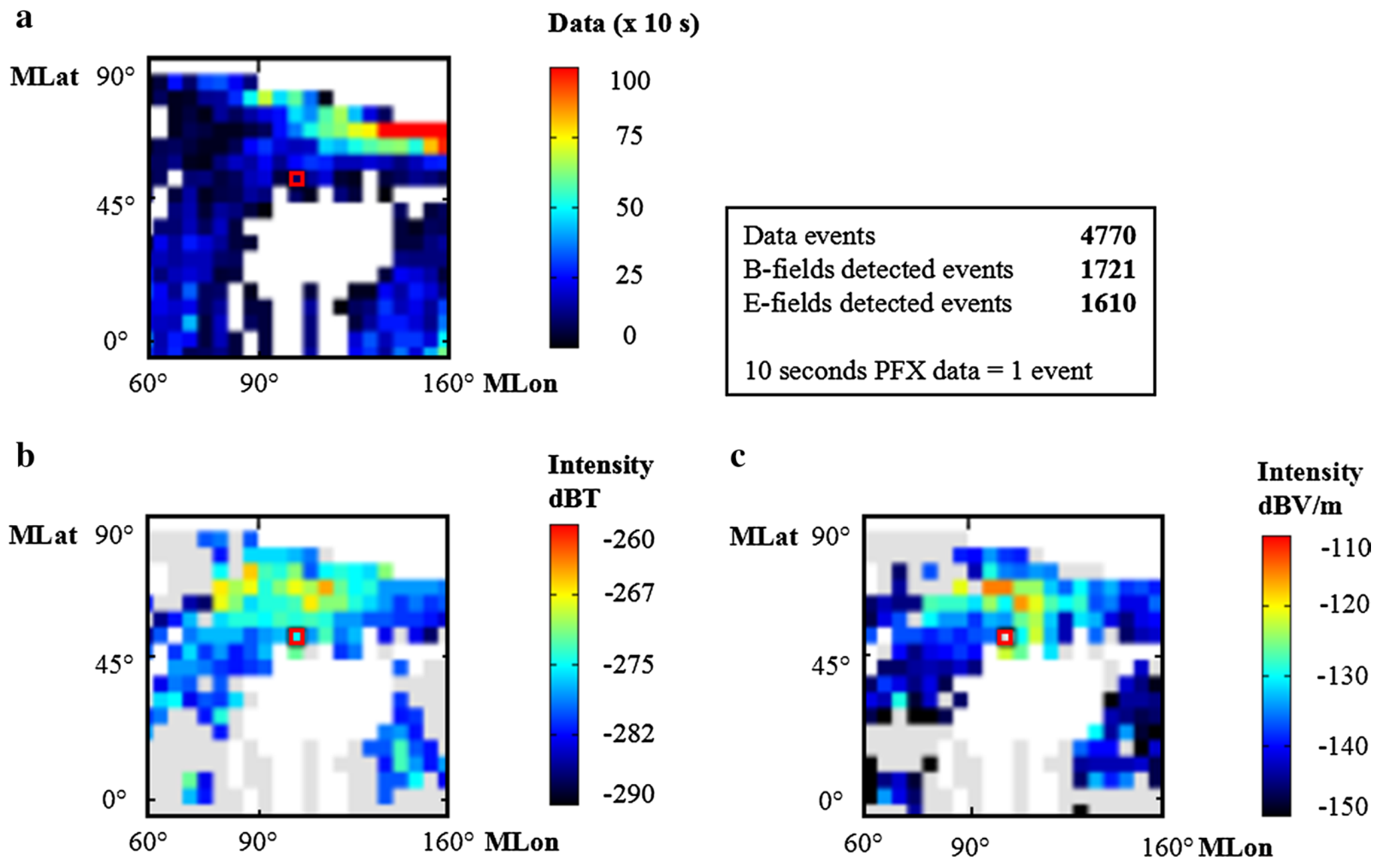

**Fig. 2** **a** Duration of data processed in the magnetic and electric field below altitude 640 km. **b** Omega signal propagation for the Norway station in the magnetic field, and **c** the electric field. The *rectangle* indicates the location of the Norway station at latitude 55.96° N and longitude 100.72° E

intensity referenced to 1 T and 1 V/m, respectively. It was found that the region where the Omega signal penetrated the ionosphere covered a radius of ~4900 km from the center of the transmitter station, between approximately latitude 0°–90° N and longitude 60°–160° E (geomagnetic coordinates), although the penetration region was not always a complete circle.

As the next step, we investigated the propagation characteristics of the Omega signal from the Norway station in a higher altitude region. In the analysis, we restricted the data measured to those between 10° to the east and 10° to the west (longitude) from the transmitter station, to illustrate the propagation of high-intensity signals within roughly 6 dB attenuation in the magnetic and electric fields compared to the center of the maximum penetration region in the meridian plane.

Figure 3 shows the Omega signal propagation pattern broadcast by the Norway station from October 1989 to September 1997 in geomagnetic coordinates. The rectangle on the map indicates the location of the station (55.96° N). We plotted the meridian map into 20 bins of altitude (631.7 km for each bin) and 36 bins of latitude (5° for each bin). The highest altitude for which we had PFX data was approximately 10,500 km, while the lowest altitude was approximately 274 km. White areas on the map indicate unavailability of PFX data, especially around the equator where $L < 2$, while more data are available at higher altitude (above 6371 km) as shown in Fig. 3a. Figure 3b, c shows the averaged intensity of the signals in the magnetic and electric fields, respectively, indicated by a color scale given at the right of each panel. The gray color in Fig. 3b, c indicates no Omega signal was detected in that region although the PFX was in operation measuring at 10.2 kHz. In Fig. 3b, c, we can clearly see that the high intensities of the Omega signal from the Norway station propagated along the Earth's magnetic field lines and that the signal reached to the opposite hemisphere (between 40° and 70° S latitude). It is also noteworthy that the Omega signal in the electric field did not attenuate as much as in the magnetic field along the magnetic

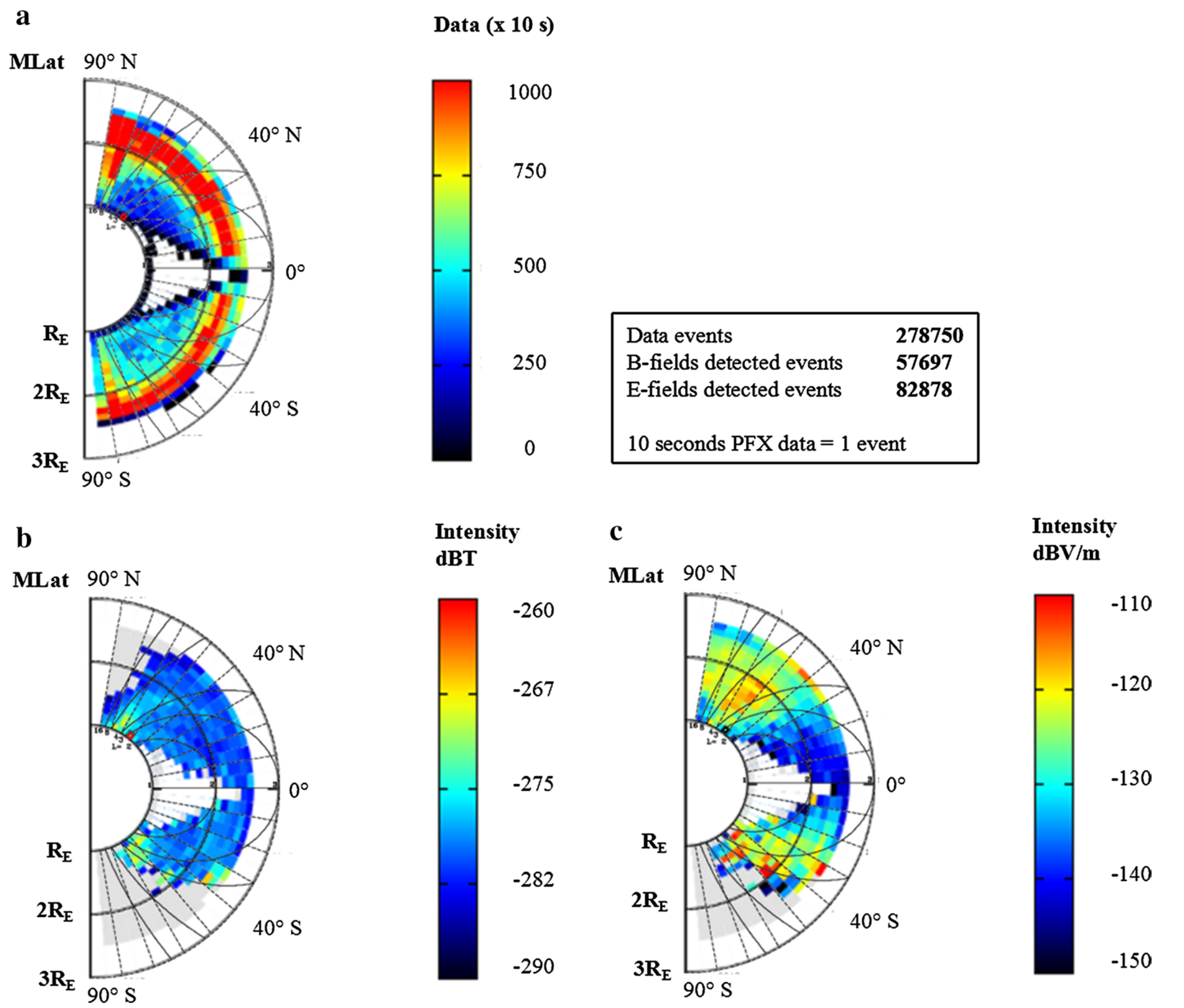

**Fig. 3** **a** Duration of data processed in the magnetic and electric field with longitude restricted to ±10°. **b** Omega signal propagation for the Norway station in the magnetic field and **c** the electric field. *Rectangle* indicates the location of the Norway station at latitude 55.96° N and longitude 100.72° E

field line at the center of the signal penetration point (~ 10° north of the transmitter station). Furthermore, the signal intensity in the electric field was distributed over a wider range in latitude (between 40° and 80° N), while the signal intensity in the magnetic field rapidly attenuated in latitudinal range.

Next, we show the propagation characteristics of the Omega signals broadcast from North Dakota. It is important to note that the stations at Norway and North Dakota were located at the same latitude (55.9° N) in geomagnetic coordinates, but at different latitudes in geographic coordinates. The Norway station was located at 56.42° N, while the North Dakota station was at 46.37° N (geographic coordinates).

Figure 4a shows the number of events used for the analysis just above the North Dakota station from October 1989 to September 1997. The station was located at latitude 55.98° N and longitude 34.83° W (geomagnetic). White areas on the map indicate unavailability of PFX data, especially around the eastern part of the transmitter station. This unavailability was caused by the tracking condition of Akebono. Over the North American region, Akebono was tracked from the tracking station at Prince Albert, Canada (Obara et al. 1990), and the North Dakota station was located southeast of Prince Albert. Because we restricted the data measured at altitude below 640 km in Fig. 4, it was not possible to track Akebono in this region. Figure 4b, c shows the averaged intensity of the signals in magnetic and electric fields, respectively, indicated by color scales given at the right of each panel. The gray color indicates no Omega signal was detected in that region. The Omega signal that penetrated the ionosphere and propagated along the Earth's magnetic field lines covered a radius of ~4700 km in the region around the transmitter station between approximately latitude 0°–90° N and longitude 80° W–20° E (geomagnetic), although the penetration region was not always a complete circle. A high-intensity area is visible just above the station transmitter at approximately −255 dBT in the magnetic field and approximately −110 dBV/m in the electric field. Compared to the Norway station, different regions of high-intensity signal are apparent, where the center of the penetration region is located at approximately latitude 55° N and longitude 34° W (geomagnetic), despite both stations being at the same latitude in geomagnetic coordinates (approximately 55.9° N).

Figure 5 shows the Omega signal propagation pattern broadcast by the North Dakota station from October 1989 to September 1997 in geomagnetic coordinates, with the same restrictions in longitude range (between 10° to the east and 10° to the west in longitude from the transmitter station) as already described in Fig. 3. The rectangle on the map indicates the location of the station (55.98° N). White areas on the map indicate unavailability of PFX data because of the orbital condition of Akebono. The gray color in Fig. 5b, c indicates that no Omega

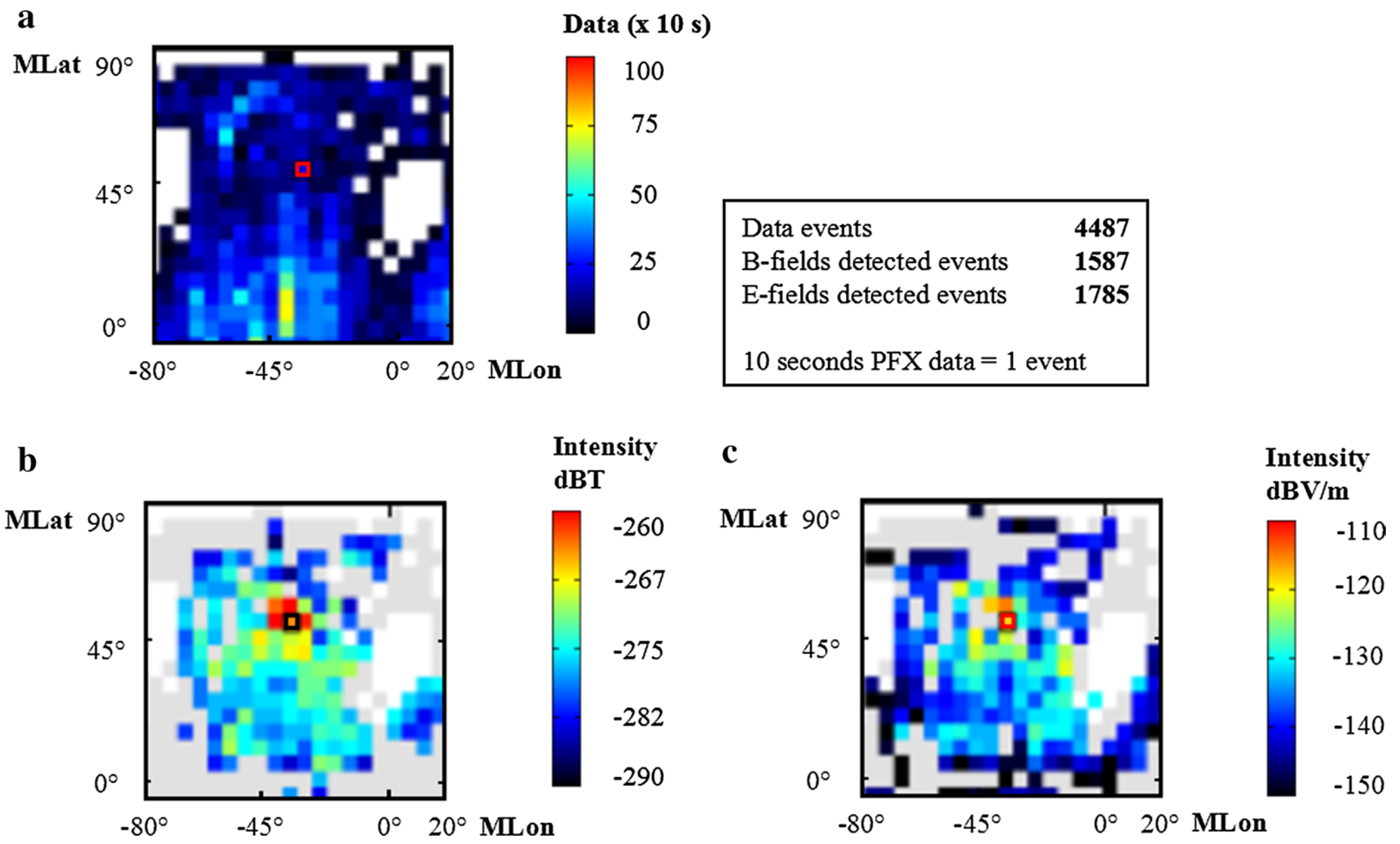

**Fig. 4** **a** Duration of data processed in the magnetic and electric fields below altitude 640 km. **b** Omega signal propagation for the North Dakota station in the magnetic field and **c** in the electric field. The *rectangle* indicates the location of the North Dakota station at latitude 55.98° N and longitude 34.83° W

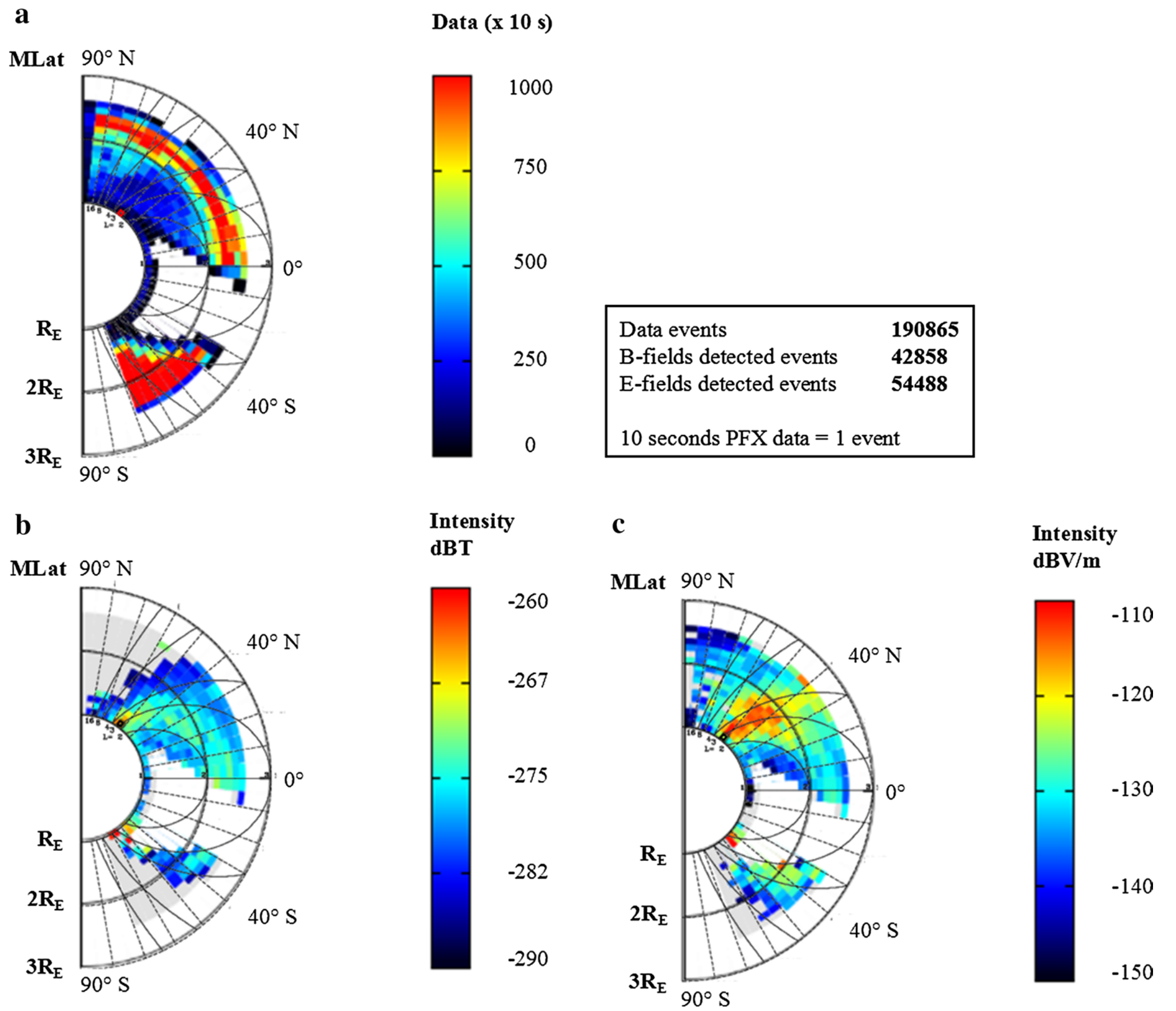

**Fig. 5** **a** Duration of data processed in the magnetic and electric fields with longitude restricted to ±10°. **b** Omega signal propagation for the North Dakota station in the magnetic field and **c** in the electric field. The *rectangle* indicates the location of the North Dakota station at latitude 55.98° N and longitude 34.83° W

signal was detected in that region. Compared to the Norway station, the high intensities of the Omega signal from the North Dakota station propagated toward the lower-latitude region between 30° and 60° along the ambient magnetic field line, and the signal propagation reached the opposite hemisphere between 35° and 65° S latitudes. Similar to the case of the Norway station, the Omega signal in the electric field did not attenuate as much as in the magnetic field along a magnetic field line up to a higher altitude region and latitude range.

## Magnetic local time dependence

As the next step, we examined magnetic local time dependence of the propagation characteristics of the Omega signals from Norway and North Dakota stations. We separated the data into eight sections referring to the magnetic local time (MLT) at the observation point every 3 h. We compared the results using the data obtained around the Norway station and the North Dakota station. Figure 6 shows the Omega signal intensity broadcast from the Norway station from October 1989 to September 1997 in geomagnetic coordinates without any altitude data restriction. The upper eight panels show magnetic field intensity, and the lower eight panels show electric field intensity. The vertical and horizontal axes indicate geomagnetic latitude and geomagnetic longitude, respectively, and the eight panels correspond to the sections separating every three MLTs. The longitude was centered at the location of the station, with a range of 100° (±50° centered at the station). The rectangle on the map indicates the location of the Norway station, which was located at latitude 55.96° N and longitude 100.72° E. The transition of the intensity level demonstrates that the nightside (MLT 21–3) has a high-intensity level at approximately −265 dBT in the magnetic field and approximately −115 dBV/m in the electric field. The dayside has the lowest intensity level of approximately −270 dBT in the magnetic field and approximately −120 dBV/m in the electric field. In addition, the dayside (MLT 9–15) clearly exhibits narrow propagation with only ~50° width of range in longitude compared to the other MLT sides.

Figure 7 shows the Omega signal intensity broadcast from the North Dakota station from October 1989 to September 1997 in geomagnetic coordinates without any altitude data restriction. The style of the panels is same as the one in Fig. 6. The rectangle on the map indicates the location of the North Dakota station, which was located at latitude 55.98° N and longitude 34.83° W. The transition of the intensity level demonstrates that the nightside had a high-intensity level at approximately −265 dBT in the magnetic field and approximately −110 dBV/m in the electric field. The dayside has the lowest intensity level of approximately −270 dBT in the magnetic field and approximately −125 dBV/m in the electric field. It is also apparent that the dayside (MLT 12–15) has narrow propagation in the magnetic field with only ~60° width of range in longitude compared to the other MLT sides. Compared with the case of the Norway station, the high-intensity regions are located around the southern part of the North Dakota station at approximately latitude 45° N and longitude 34° W, although there is a similar transition in the propagation pattern between each MLT and the signals tend to propagate further on the nightside (MLT 21–3). If compared to the signals from Norway, the signals from North Dakota on nightside show higher level of intensity (about 5 dB) in the electric field and about 5 dB in the magnetic field and were widely distributed. It also shows a lower level of intensity (approximately 5 dB) in the magnetic and electric fields and a narrower region on dayside.

## Seasonal variation

In this section, we demonstrate the results of seasonal variation of the Omega signal propagation. In the analyses, we separated the datasets obtained over eight years into four seasons approximately centered at the equinox and solstice of each year. The data range for each season is separated as follows: from February 1 to April 30 as "March equinox," from May 1 to July 31 as "June solstice," from August 1 to October 31 as "September equinox," and from November 1 to January 31 as "December solstice."

Figure 8 shows the spatial distribution of the Omega signals broadcast by the Norway station from October 1989 to September 1997. The upper four panels show magnetic intensity, and the lower four panels show electric intensity. As shown in Fig. 3, Omega signals tend to be refracted by the geomagnetic field lines and propagate from northern hemisphere to southern hemisphere. In order to evaluate their deviation and/or broadening from the geomagnetic field lines, we plotted the intensity map taking its vertical axis as invariant latitude (ILAT) of the observation point and the horizontal axis as geomagnetic longitude. The rectangle on the map indicates the location of the station. It was found that the high-intensity

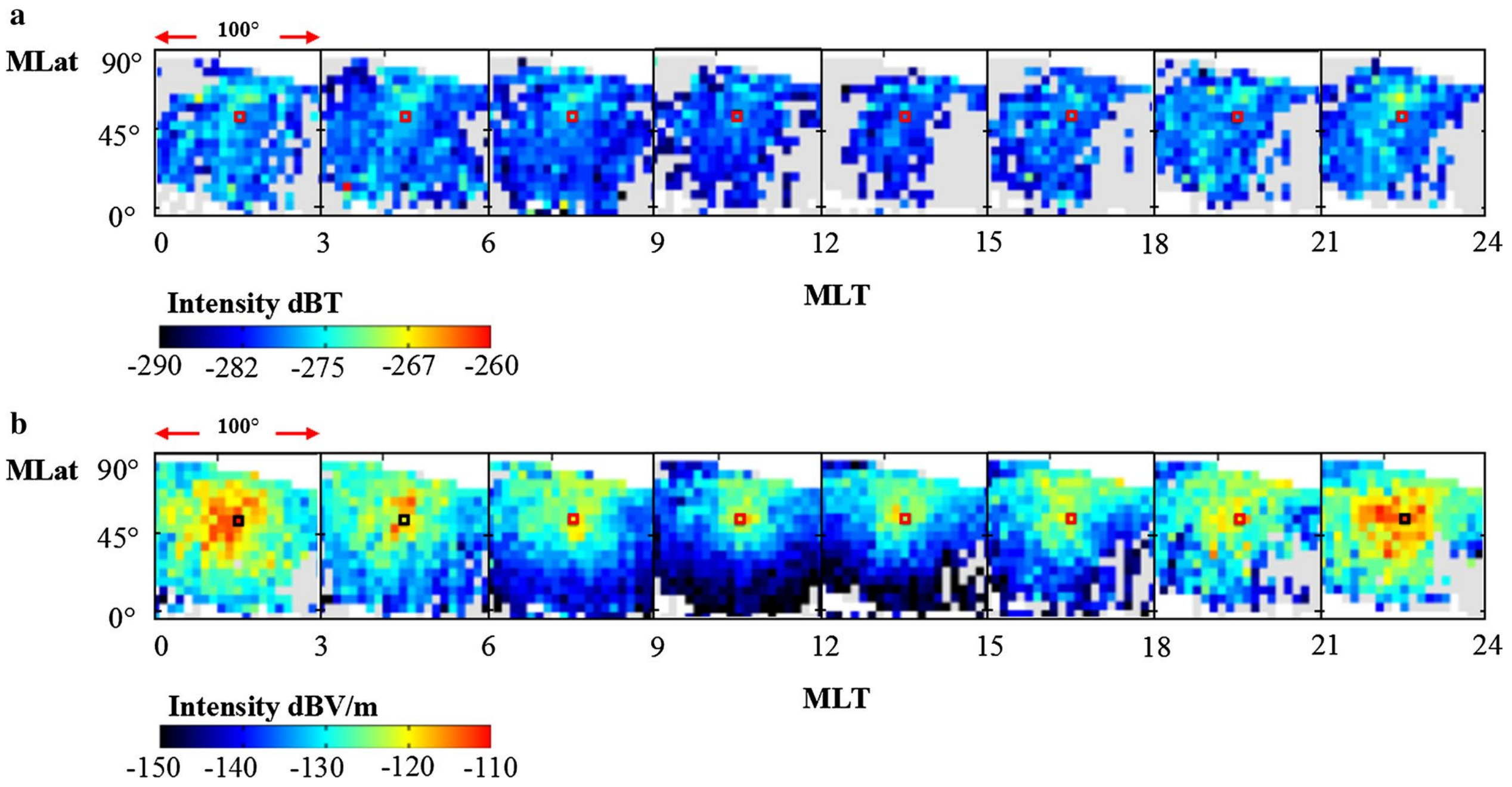

**Fig. 6** **a** Propagation of the Omega signal from the Norway station based on local time in the magnetic field and **b** in the electric field. The *red arrow* indicates the width of the map (~100°) with location of the station in the center (indicated by the *rectangle*)

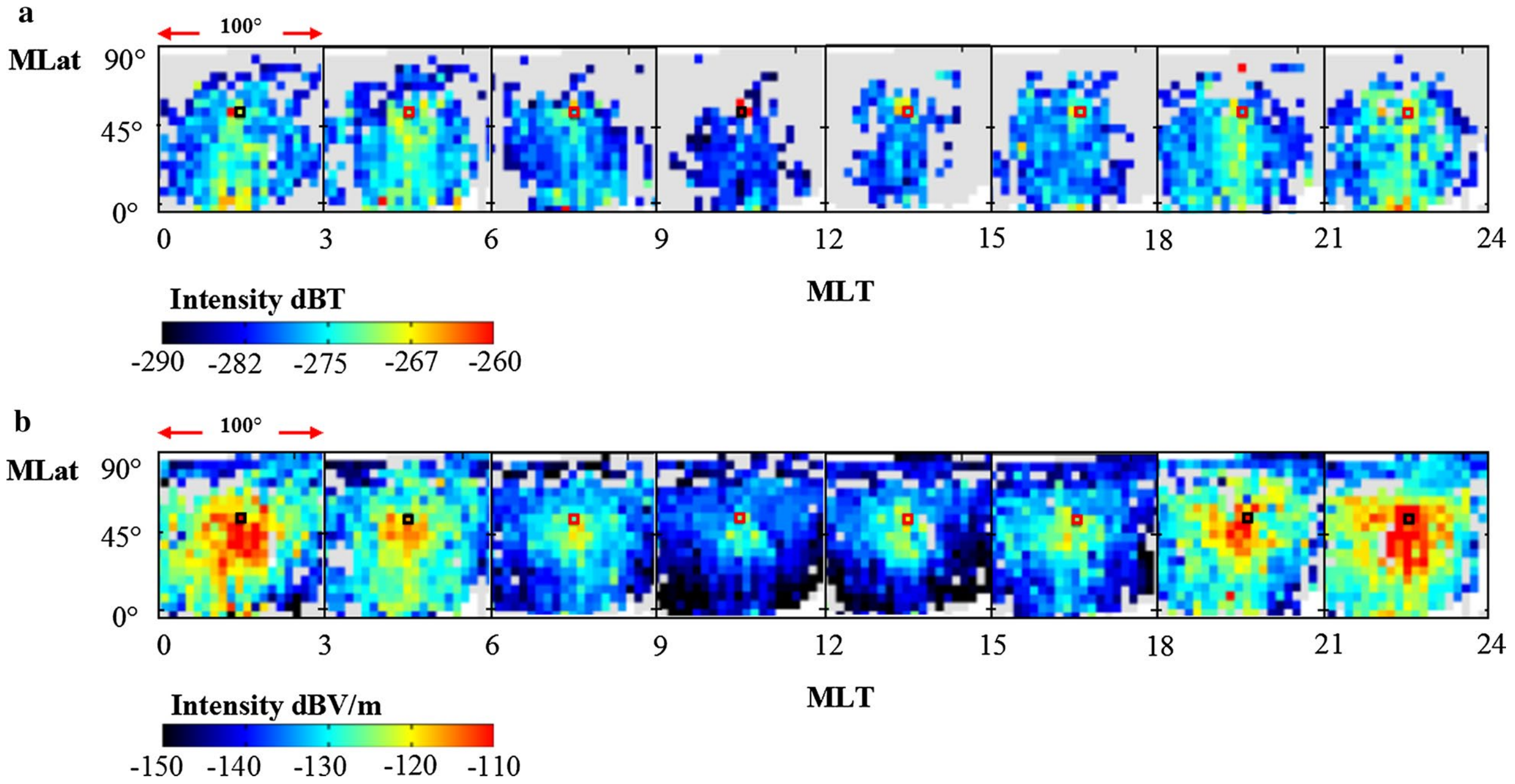

**Fig. 7** **a** Propagation of the Omega signal from the North Dakota station based on local time in the magnetic field and **b** in the electric field: the *red arrow* indicates the width of the map (~100°) with the station in the center (indicated by the *rectangle*)

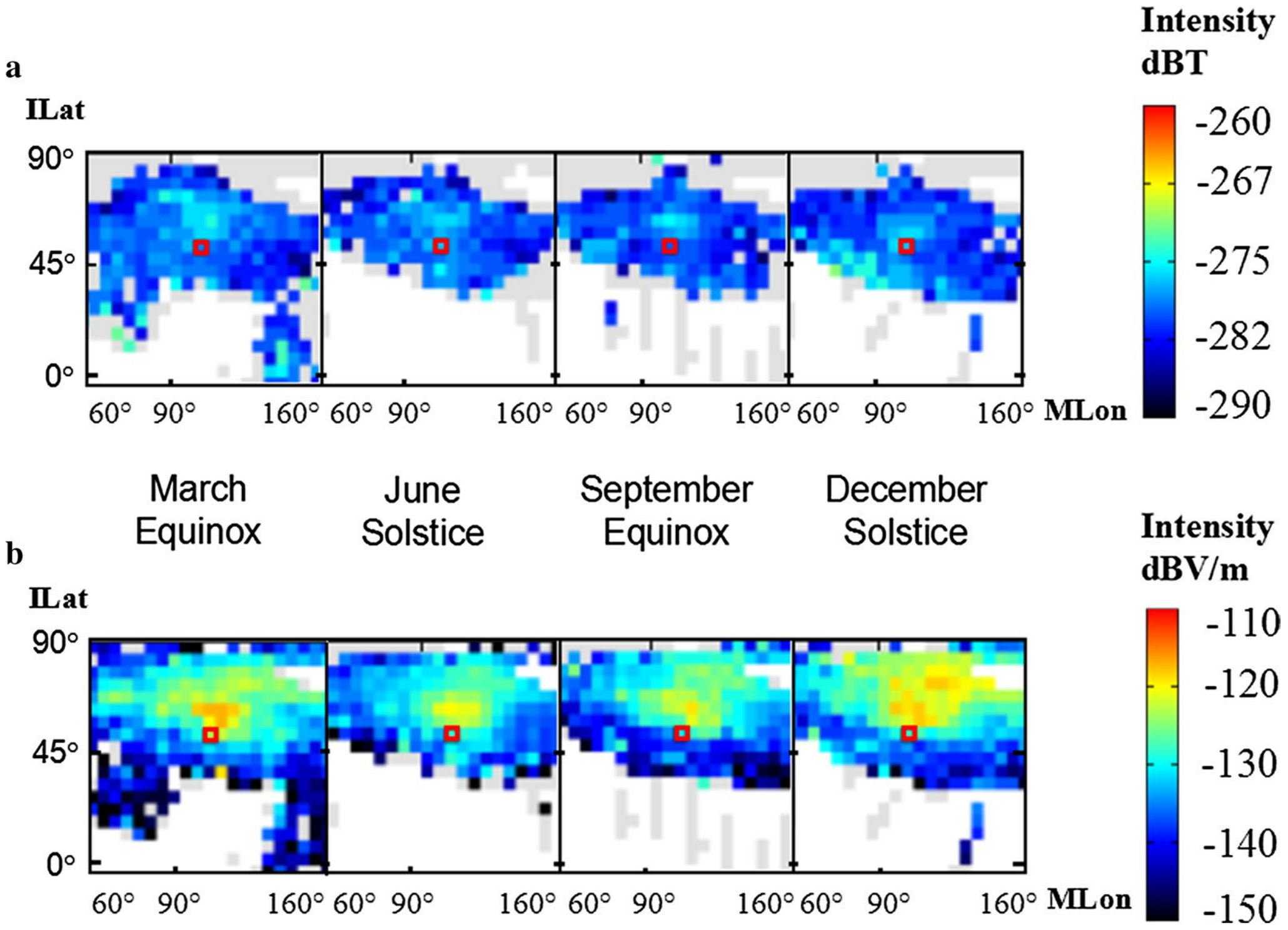

**Fig. 8** **a** Seasonal propagation of the Omega signal from the Norway station in the magnetic field and **b** in the electric field based on invariant latitude

region deviated 10° to higher invariant latitude from the transmitter. Intensities in the high-intensity region were approximately −275 dBT for the magnetic field and approximately −117 dBV/m for the electric field. Intense signals were detected in a wider region in December solstice, while the high-intensity region became the smallest in June solstice. In addition, it is notable that the high-intensity region in the electric field was located only at 10° higher invariant latitude from the transmitter around the time of the June solstice. However, it gradually spread northward (higher latitude) in September equinox and became the widest in December solstice. The region shrunk southward again in March equinox.

To understand the attenuation ratio between magnetic field and electric field based on the propagation distance from the Earth's surface, we show the magnetic and electric intensity distributions along the magnetic field lines in Fig. 9a, b, respectively. The vertical axis indicates invariant latitude, and the horizontal axis indicates distance from the Earth's surface along the magnetic field lines, with maximum range of 12,742 km. Small red lines indicate the latitude of the station location.

We clearly see that the magnetic intensity attenuated gradually as the propagation distance became larger, while the electric intensity did not attenuate much even in the middle at distance at ~5700 km. In addition, signals tend to propagate within the narrower invariant latitude region in June solstice, but they propagate in a wider invariant latitude region, especially toward the higher-latitude region in December solstice and March equinox.

Figure 10 shows the spatial distribution of Omega signals broadcast by the North Dakota station from October 1989 to September 1997. The vertical axis indicates the invariant latitude (ILAT) of the observation point, and the horizontal axis indicates geomagnetic longitude. The rectangle on the map indicates the location of the station. We can also find that the high-intensity region of the propagation deviated ~ 10° to higher latitude from the transmitter, as in the Norway case (Fig. 8). Intensities in the high-intensity region were approximately −270 dBT for the magnetic field and approximately −115 dBV/m for the electric field. Compared with the Norway case shown in Fig. 8, however, it is obvious that the high-intensity region was almost constant throughout the

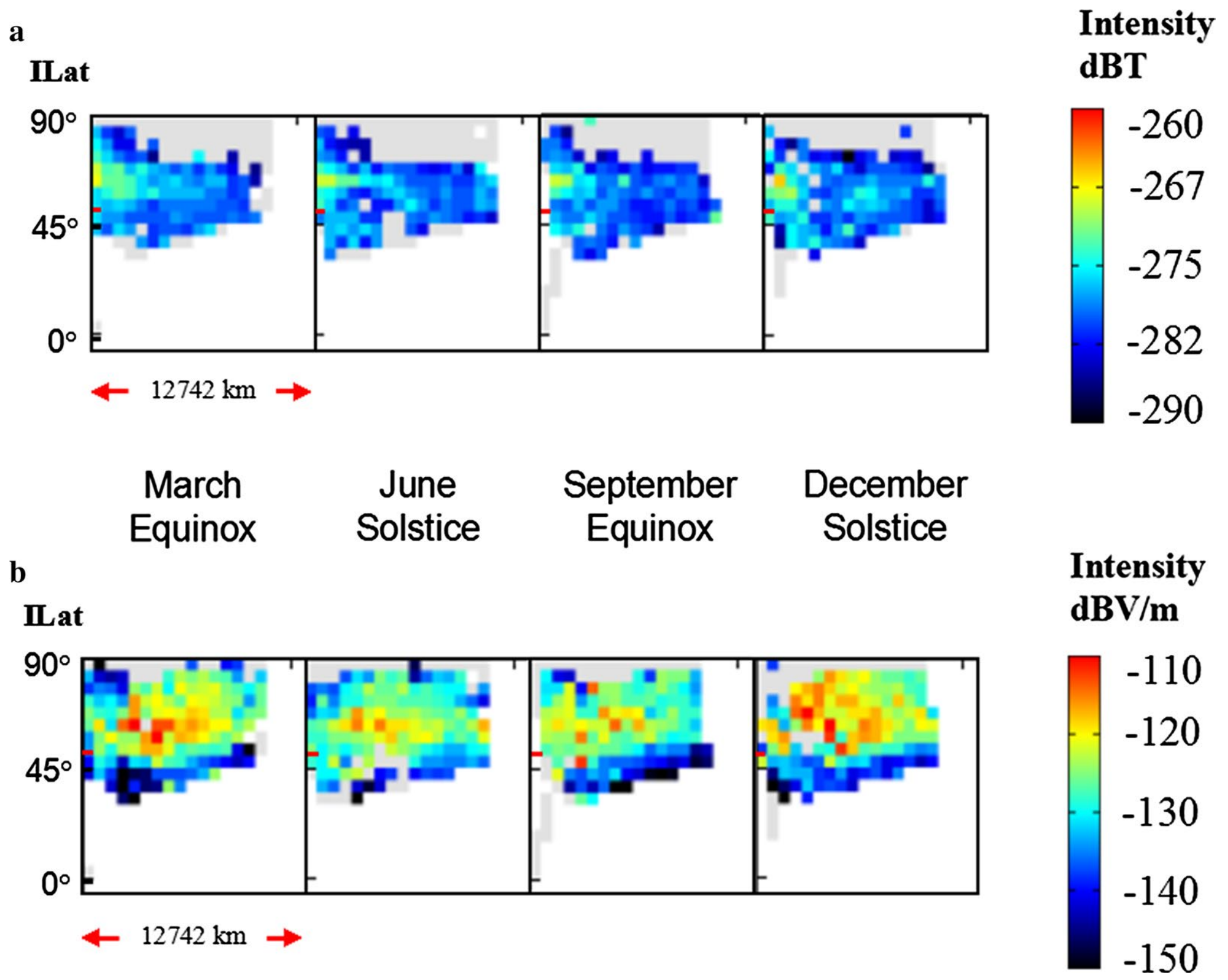

**Fig. 9** **a** Seasonal propagation of the Omega signal from the Norway station in the magnetic field and **b** in the electric field based on invariant latitude and the propagation distance from the Earth's surface

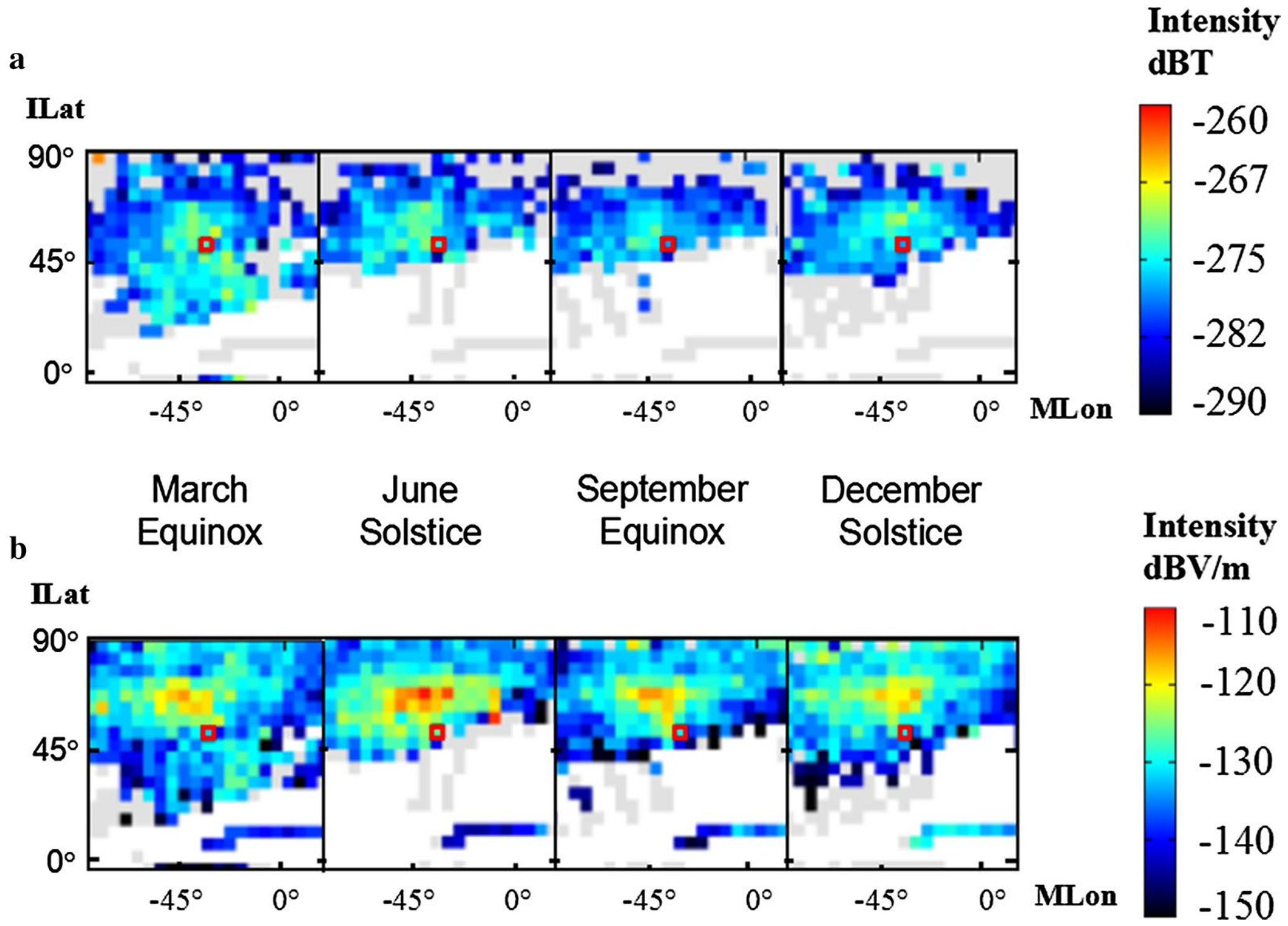

**Fig. 10** **a** Seasonal propagation of the Omega signal from the North Dakota station in the magnetic field and **b** in the electric field based on invariant latitude

year and was not widely distributed even in December solstice.

Figure 11a, b shows the magnetic and electric intensity distributions of the Omega signals broadcast by the North Dakota station along the magnetic field lines. We also find that the Omega signals from North Dakota tended to propagate along the ambient magnetic field lines within a limited, invariant latitude. We assume that this happened because of the different geographic latitudes even though the stations have the same geomagnetic latitude. That is, the Norway station was located at higher geographic latitude, and thus, Omega signals could propagate widely because of the longer night in December solstice.

## Annual variation

Finally, we studied the annual variation of the propagation characteristics to evaluate the effects of solar activity. We separated the data and illustrated the results for each year. As shown in Fig. 1 and Table 1, there is no significant difference in the dataset coverage except for 1989, in which the available data were obtained for only three months. In addition, we selected the data measured at an altitude below 8200 km for the entire year, because the satellite's altitude decreased every year and no data are available above that altitude in 1997. Figure 12 shows the statistical results of the Omega signal propagation pattern broadcast by the Norway station from October 1989 to September 1997 in geomagnetic coordinates. The vertical axis indicates geomagnetic latitude, and horizontal axis indicates geomagnetic longitude within a range of 100° (±50° centered at the station), and nine panels corresponding to 1989–1997 are shown. The rectangle on the map indicates the location of the station. It was found that the intensity level was low between 1989 and 1992, while the average intensity level was higher, and the propagation region of the Omega signal was wider, from 1993 to 1997. Based on sunspot cycle data (NASA/Marshall Solar Physics 2016; Usoskin 2008), the number of sunspots on the surface of the sun increased between 1989 and 1992 during the solar maximum period, while it decreased between 1993 and 1997 corresponding to the solar minimum period. We demonstrate that solar activity significantly affected the propagation and penetration of the Omega signal in the ionosphere. The Omega signal propagation intensity near the station during the solar maximum (1991) was approximately −280 dBT in the magnetic field and approximately −120 dBV/m in the electric field. During the solar minimum (1996), higher-intensity signals that propagated wider and farther from the transmitter station are apparent, approximately −270 dBT in the magnetic field and approximately −115 dBV/m in the electric field. Based on these maps, the coverage radius and intensity strength of the Omega signal that penetrated and propagated in the ionosphere changed by a few thousand kilometers every year.

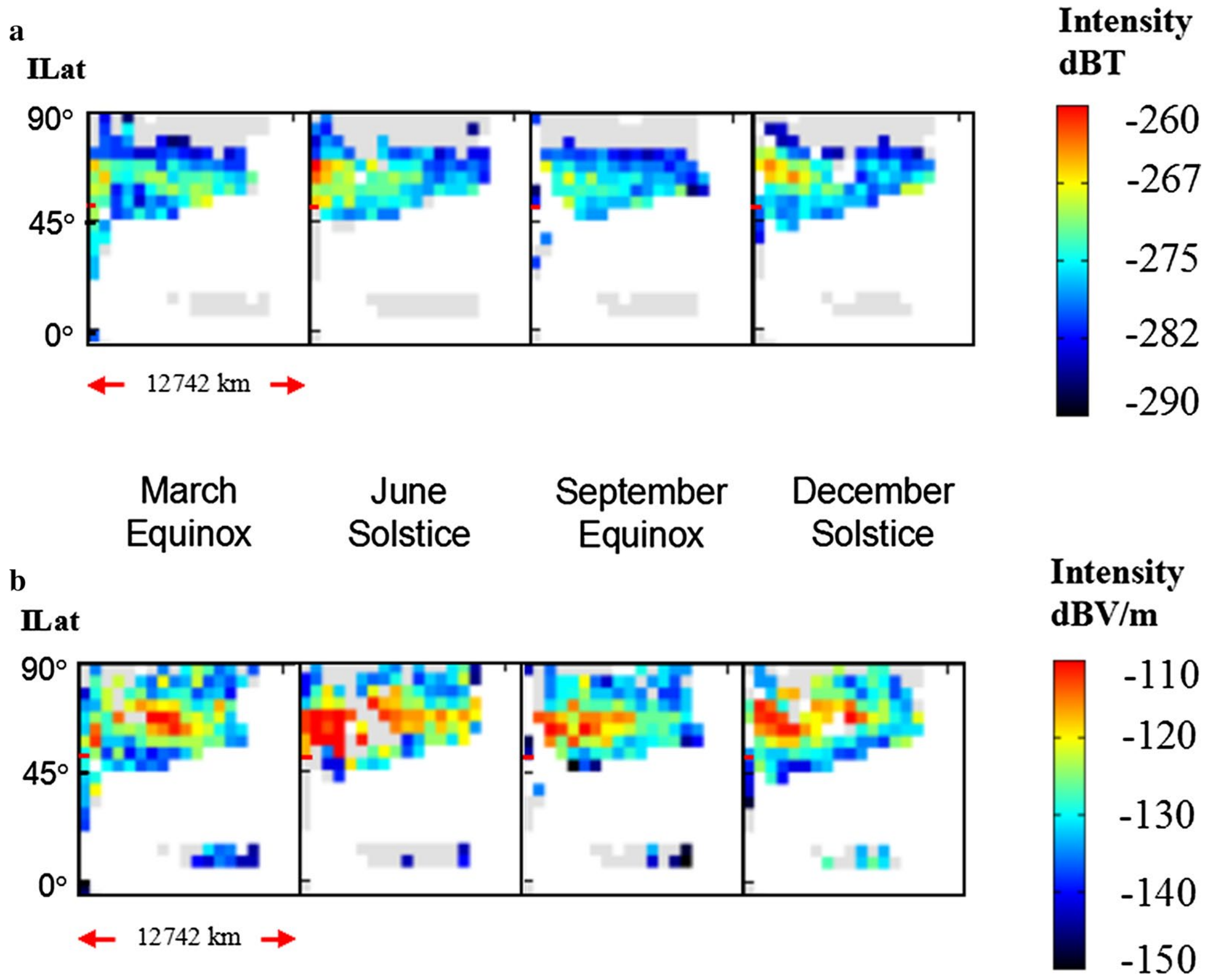

**Fig. 11** **a** Seasonal propagation of the Omega signal from the North Dakota station in the magnetic field and **b** in the electric field based on invariant latitude and the propagation distance from the Earth's surface

## Discussion and conclusions

We studied the high–middle-latitude Omega stations located in Norway and North Dakota and evaluated how widely and how far the Omega signals penetrated through the ionosphere over the station. Our analysis revealed that, for stations located at almost the same latitude in geomagnetic coordinates, the Omega signal propagated differently along the Earth's magnetic field. Omega signals broadcast from both stations tended to propagate along the geomagnetic field lines and reached the opposite hemisphere. However, the signals from Norway propagated in the higher-latitude region compared to the ones from North Dakota. This phenomenon was a result of the stations being at different geographic coordinate latitudes: The Norway station was located at 56.42° N and the North Dakota station at 46.37° N. We suggest that effect of the attenuation ratio in the ionosphere worked differently because of this difference. According to the study by Clilverd et al. (2008), the VLF wave of higher $L$-shell ($L > 1.5$) is highly ducted. We suggest that this fact might have affected the difference in propagation between Norway and North Dakota. However, we need to investigate further by examining the wave normal directions of Omega signals from both stations using the PFX data statistically. In the present study, we mainly focus on the magnetic and electric intensities in the northern hemisphere to evaluate how much the Omega signals penetrated through the ionosphere over the station. However, we also demonstrate that the signals tend to be refracted by magnetic field lines and propagate toward the opposite hemisphere. Further investigation of the conjugate region will be performed in future work.

Next, we examined the transmittance of the Omega signals as a function of magnetic local time to evaluate the attenuation effect of the ionosphere and plasmasphere. We found that the Omega signal tended to propagate farther on the nightside, where the electron density

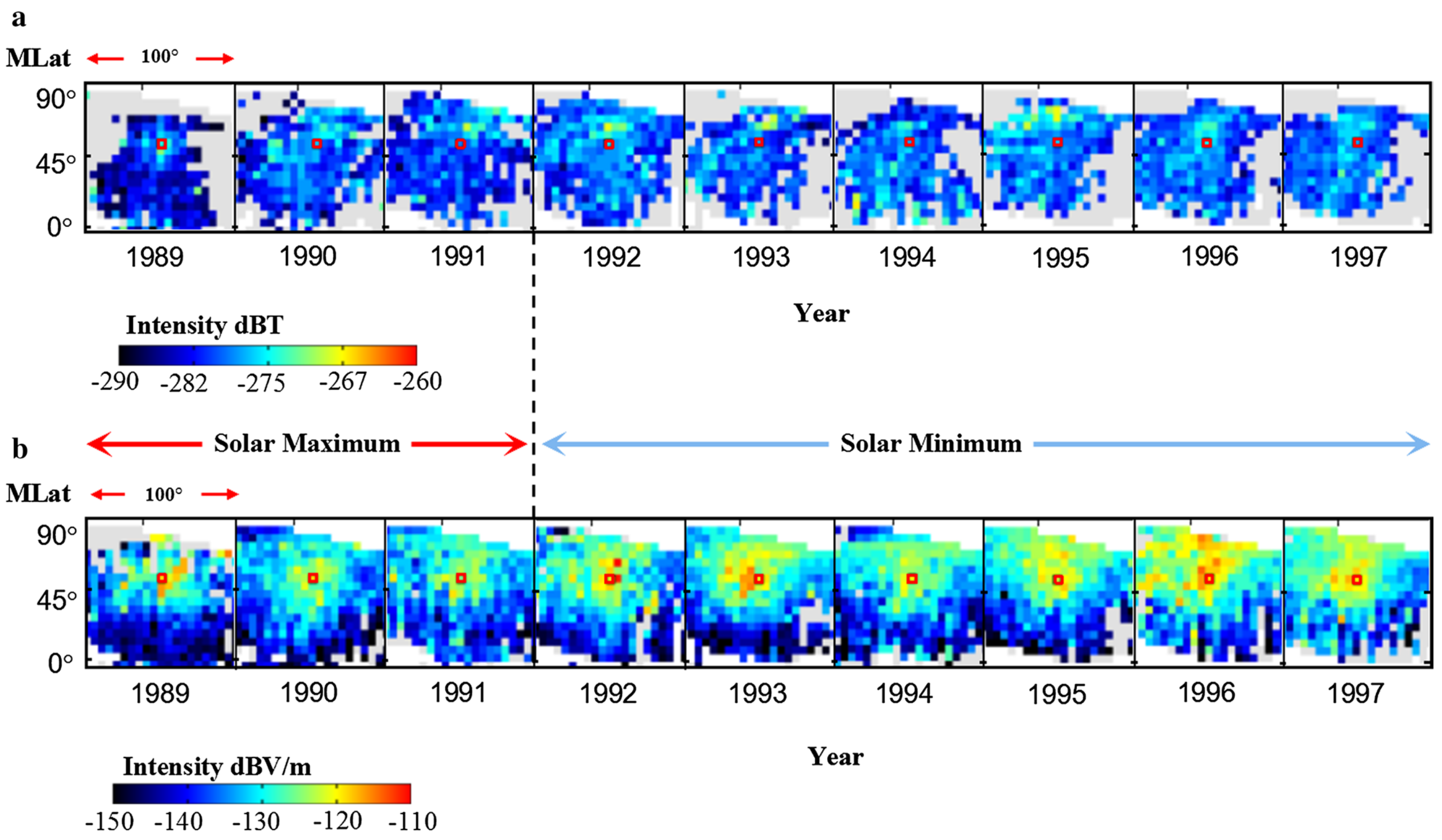

**Fig. 12** Annual propagation of the Omega signal from the Norway station in the magnetic field and **b** in the electric field. The *red arrow* indicates the width of the map (~100°) with location of the station in the center (indicated by the *rectangle*)

in the ionosphere was lower, than on the dayside. This result is consistent with previous studies by Collier et al. (2006) and Oike et al. (2014), in which they studied the occurrence frequencies of whistlers. Collier et al. (2006) demonstrated that the detection frequency of whistlers on the ground was higher at night than at daytime. On the other hand, Oike et al. (2014) studied the whistlers detected in the plasmasphere and showed that the occurrence frequency was maximum after sunset. They suggested that this was caused by the combination of the frequency of occurrence of lightning and the attenuation effect of the ionosphere; that is, the occurrence frequency of lightning peaked in late afternoon but the attenuation ratio was smaller at night, so the detection peak on the spacecraft reached maximum after sunset. In the case of the ground measurement, it is necessary to penetrate the ionosphere twice, and thus, whistlers tend to be detected at night on the ground. In that sense, our statistical study is valuable for quantitatively evaluating the attenuation ratio of VLF waves in the ionosphere.

As for the propagation characteristics in the plasmasphere, we showed that the signal intensity in the magnetic field gradually decreased along the propagation path, while the attenuation ratio in electric field was lower; the signals could propagate more widely and farther (to the other hemisphere) from the original transmission station.

We also studied the seasonal variation of Omega signals to evaluate the trend of propagation characteristics. We found that intense signals from the Norway station were detected in a wider region in December solstice, while the high-intensity region became the smallest in June solstice. However, the signals from North Dakota were not widely distributed and tended to propagate in a narrower region along the ambient magnetic field. This is probably because the Norway station was located at higher geographic latitude, and thus, the night was longer than at the North Dakota station during the December solstice. Thus, the signals from the Norway station were not attenuated at that time and propagated widely in the higher-latitude region.

Finally, we studied the annual variation of the propagation characteristics of the Omega signals to evaluate the effects of solar activity. We found that in 1991, when solar activity was at a maximum, the Omega signal propagated at a lower intensity. In contrast, when solar activity was at minimum in 1996, the Omega signal propagated at a higher intensity and reached farther from the transmitter station. We assumed that plasmaspheric electron density and temperature affected the propagation of the Omega signal.

As described in the introduction, quantitative study of Omega signals using eight years of PFX data from October 1989 to September 1997 is quite important to evaluate the attenuation ratio of VLF waves propagating in the ionosphere and plasmasphere as a function of magnetic local time, season, and annual solar activity. In the future, extensive studies of the delay time and wave normal direction are necessary to quantitatively clarify the global features of the ionosphere and plasmasphere.

#### Abbreviations

IC: integrated circuit; CDF: common data format; PFX: Poynting flux analyzer; VLF: very low frequency; WIDA: wide dynamic range amplifier; MLT: magnetic local time.

#### Authors' contributions

IMADS carried out the study and drafted and completed the manuscript. YK participated in its design, coordination, and revision. YG reviewed the resulting paper and the revision. All authors read and approved the final manuscript.

#### Author details

[1] Graduate School of Natural Science and Technology, Kanazawa University, Kakuma-machi, Kanazawa 920-1192, Japan. [2] Present Address: Udayana University, Badung, Bali 80361, Indonesia.

#### Acknowledgements

The data used in this research were provided by the Akebono Project which was conducted by JAXA, Japan.

#### Competing interests

The authors declare that they have no competing interests.

#### Availability of data and materials

The data and materials used in this research are available on request basis to the corresponding author, Prof. Yoshiya Kasahara (kasahara@is.t.kanazawa-u.ac.jp).

#### Funding

This research was partially supported by a Grant-in-Aid for Scientific Research from the Japan Society for the Promotion of Science (#16H01172 and #16H04056).

## Publisher's Note